# Structural, Transport and Magnetic Properties of $La_{0.67}Ca_{0.33}Mn_{1-x}Ta_xO_3$

L. Seetha Lakshmi[a,b,@], K. Dörr[b], K. Nenkov[b], V. Sridharan[a], V. Sankara Sastry[a] and K. -H. Müller[b]

[a]Materials Science Division, Indira Gandhi Centre For Atomic Research, Kalpakkam, 603102, India

[b]Institute of Metallic Materials, IFW Dresden, Postfach 270116, Dresden 01171, Germany

**Abstract**

For the first time, the effect of $Ta^{5+}$ substitution at the Mn site of $La_{0.67}Ca_{0.33}MnO_3$ compound is reported. A significant increase in lattice parameters and unit cell expansion indicate larger structural modification upon substitution. The ferromagnetic-metallic ground state is modified to a cluster glass insulator for x>0.03. The observed suppression of the ferromagnetic transition temperature $T_C$ of ~ 39K/at.% might be the highest reported in Mn-site substituted $La_{0.67}Ca_{0.33}MnO_3$. The well-known effects of changed carrier density, larger average ionic radius at Mn site and dilution of Mn sublattice appear insufficient to explain this strong suppression of $T_C$.

## 1. Introduction

In manganites, the subtle balance between charge, spin, lattice and orbital degrees of freedom leads to a variety of phases with different physical properties. Hence the substitution studies at the Mn site offer an extremely useful handle on the properties of the system. Numerous investigations [1-5] carried out on Mn site substitutions show that all the substituents in ferromagnetic metallic manganites, irrespective of their electronic and magnetic nature, lower the ferromagnetic transition temperature, but to different extents. At a certain higher level of substitution, ground state is found to modify to a cluster glass in some cases [1,2,6,7]. The decrease in transition temperature is generally attributed to the weakening of double-exchange (DE) interaction strength. In our previous studies on Mn site substitutions with both diamagnetic and paramagnetic ions [8-10], we rationalized the variation of extent of suppression in the transition temperature with concentration (dTc/dx) in terms of local structural modification due to size mismatch and local magnetic coupling between the magnetic moments of substituents and Mn ion. In this paper, we address the role of yet another important parameter, the valence state of the substituent in modifying the ground state properties. For the first time, we report the effect of pentavalent Ta substitution at the Mn site of $La_{0.67}Ca_{0.33}MnO_3$. By virtue of its closed shell configuration, $Ta^{5+}$ ion does not introduce any additional magnetic coupling. From our present studies, we show that charge state (valence) of the substitution has a dominant role in modifying the ground state of colossal magnetoresistive manganites.

## 2. Experiment:

Polycrystalline samples were synthesized by the standard solid state reaction method. Final sintering of the samples at 1500°C for 24 hrs in air was carried out in a single batch to ensure identical sintering conditions. The density of the sintered samples is about 95 % of the ideal value. The high statistics room temperature powder XRD patterns were recorded in the 2θ range, 15-120° using CuKα radiation (STOE). The lattice parameters and unit cell

---

[@] Corresponding Author: Email : slaxmi73@hotmail.com, seetha@igcar.ernet.in

volume were determined using the GSAS Rietveld refinement program [11] and SPUDS program [12] to generate a starting model for the Rietveld analysis. The resistivity measurements ($\rho(T)$) in zero field and 7T were performed on rectangular bar shaped pellets by conventional four point method. A superconducting split coil magnet was employed for steady magnetic fields up to 7T with magnetic field applied parallel to the current. The temperature variation of ac susceptibility ($\chi'(T)$) with ac field of 1 Oe and at a frequency of 133 Hz was measured using a Lakeshore 7000 Series susceptometer. The para-ferromagnetic transition temperature ($T_C$) was determined by tangent method and corresponds to the onset of $\chi'$ signal.

## 3. Results and discussion

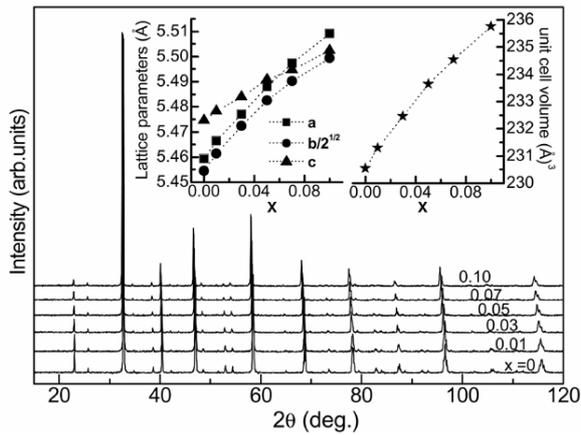

**Fig:1**. High statistics room temperature powder XRD patterns of $La_{0.67}Ca_{0.33}Mn_{1-x}Ta_xO_3$ ($0 \leq x \leq 0.10$) compounds. Inset shows the variation of lattice parameters (a, b and c) and unit cell volume with x.

Powder XRD patterns indicate that the compounds are monophasic with orthorhombic structure and Pnma symmetry (Fig.1). The lattice parameters a, b and c show a linear increase with Ta substitution, hence an overall linear expansion of the unit cell is observed (Fig.1 inset). More detailed analysis of the crystal structure will be published elsewhere [13]. The compounds with $x \leq 0.03$ exhibit a metal to insulator transition (MIT) at $T = T_{MI}$ in $\rho(T)$ (Fig.2). Close to $T_{MI}$, the para to ferromagnetic transition is observed at $T = T_C$ (Fig.3). Beyond x =

0.03, $\rho(T)$ curves exhibit an insulating behaviour in the entire temperature range of study. (The experimental set up limits the resistivity measurements to about $10^6$ $\Omega$.cm.) In particular, no anomaly is detected near $T_c$, indicating a non-metallic nature of the ferromagnetic phase. It is noteworthy that both the $T_{MI}$ and $T_c$ reduction with increasing x are the largest reported for Mn site substituted $La_{0.67}Ca_{0.33}MnO_3$ [1-5,8-10]. While $T_{MI}$ decreases at a constant rate of $\sim$ 36 K/ at.%, a linear decrease of similar rate ($\sim$39 K/ at.%) is observed for $T_c$ up to x = 0.03, which levels off to a much smaller rate (10 K/ at.%) for higher Ta concentrations (inset of Fig.3).

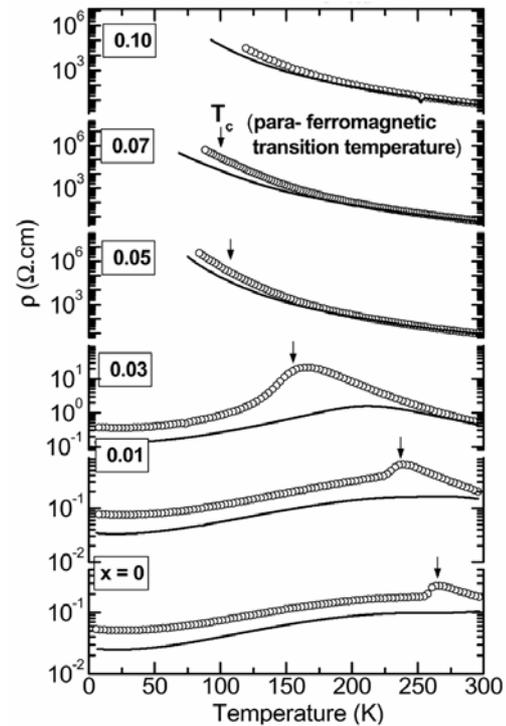

**Fig:2**. Variation of resistivity of $La_{0.67}Ca_{0.33}Mn_{1-x}Ta_xO_3$ ($0 \leq x \leq 0.10$) with temperature in the absence of magnetic field (O) and at a magnetic field of 7 T (—).

Under an applied magnetic field (H) of 7T the compounds with $x \leq 0.03$ exhibit a significant reduction in $\rho$ with a shift of $T_{MI}$ to higher temperature, characteristic of CMR manganites. No field induced metallic state could be found for $x \geq 0.05$ at low temperatures. Interestingly, for $x \geq 0.05$, $\chi'(T)$ shows a cusp like feature below $T_c$, followed by a broad shoulder at lower temperatures (fig.3). This shoulder develops into a separate peak if anappropriate small dc magnetic field is applied [13]. It is attributed to freezing of magnetic clusters. Other typical features of a glassy state have been observed.
i) the irreversibility between zero-field-cooled (ZFC) and field-cooled (FC) magnetization with a pronounced drop of ZFC magnetization at low temperatures and reduced irreversibility with enhanced measuring field, and (ii) slow relaxation of magnetization below 50 K.

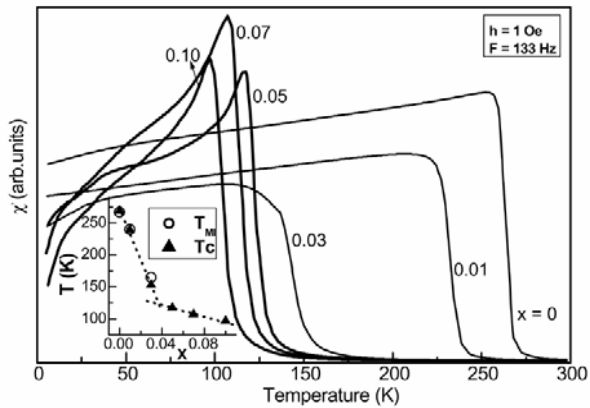

**Fig:3**. Variation of ac susceptibility of $La_{0.67}Ca_{0.33}Mn_{1-x}Ta_xO_3$ ($0 \leq x \leq 0.10$) with temperature. Inset: Variation of transition temperatures, $T_{MI}$ (○) and $T_C$ (▲) as a function of x; lines are a guides to the eye.

In the following, we consider the microscopic consequences of $Ta^{5+}$ substitution. Taking charge neutrality into account, pentavalent Ta is expected to strongly shift the average valence state of Mn towards 3+, according to $La^{3+}_{0.67}Ca^{2+}_{0.33}Mn^{3+}_{(0.67+x)}Mn^{4+}_{(0.33-2x)}Ta^{5+}_xO_3$. The ionic radii (IR) used to estimate the change of average ionic radius are 0.645 Å ($Mn^{3+}$), 0.640Å ($Ta^{5+}$), 0.53Å ($Mn^{4+}$), 1.216 Å ($La^{3+}$) and 1.18Å ($Ca^{2+}$) [14]. Thus, the appreciable increase of the lattice parameters with $Ta^{5+}$ substitution (Fig.1) can be understood as a result of the larger average ionic radius at the Mn site, though electron-lattice coupling might give an additional influence. Further, there is a substantial drop of the carrier density ($Mn^{4+}$ concentration) with increasing x. Indeed, the consequences of these two effects (reduced carrier density and increased ionic radius at the Mn site) should be displayed in similar way by compounds $La_{1-z}Ca_zMnO_3$ [15] if the same level of doping ($Mn^{4+}$ content) is compared. The Mn site ionic radius is nearly equal for respective compounds, while the La site ionic radius remains nearly constant due to similar ionic radius of La and Ca. However, the Ta substituted samples exhibit much lower values of $T_c$. For instance for x = 0.05, 23 % of Mn sites contain $Mn^{4+}$ ions (or 21.9 % of present Mn ions are $Mn^{4+}$), leading to $T_c$ = 117 K and low temperature glassy behavior. On the other hand, ferromagnetism and metallicity are found in $La_{0.77}Ca_{0.23}MnO_3$ ($T_C$~220 K) with same doping level [15]. Therefore, the effects of carrier density and ionic size seem insufficient to account for the observed strong suppression of ferromagnetism. Larger structural modification involving an increase in the average Mn-O bond length ($d_{Mn-O}$) and a substantial decrease in Mn-O-Mn bond angles (<Mn-O-Mn>) has been detected for Ta substituted compounds [13] which is expected to decrease the DE interaction strength. The diamagnetic $Ta^{5+}$ ion ($4d^{10}$), by virtue of its closed shell configuration, just dilutes the magnetic network, and dilution generally results in a reduced magnetic ordering temperature. However, the degree of dilution is low in the studied compounds, and other diamagnetic substitutions for Mn affect $T_c$ much less than Ta. Hence, another microscopic effect seems necessary to understand the experimental result. As pointed out by Alonso *et. al.* [17], the changes in charge state at the Mn site may create a random local electrostatic potential that attracts or repels charge carriers (holes). Taking the local electrostatic potential into account, calculations of these authors (carried out for $Ga^{3+}$ ions) indicate enhanced suppression of $T_C$ and glassy states. Due to the large positive charge of $Ta^{5+}$ ion, one would expect a strong electrostatic effect for this substitution, though the potential is positive in this case. We suggest that this mechanism might contribute to the observed unusually strong suppression of the itinerant DE ferromagnetism. In the presence of antiferromagnetic superexchange interactions and spatial disorder, the appearance of glassy behavior is not unusual in manganites. The ability of $Ta^{5+}$ ion to induce cluster glass and insulating behavior at very low Ta concentration (x ~ 0.05) if compared to other trivalent and tetravalent substitutions for Mn underlines the significance of charge state (valence) of the substituent in modifying the ferromagnetic-metallic ground state of CMR manganites.

**Acknowledgement**

Support by DFG, FOR520 is gratefully acknowledged. LSL also thanks CSIR, India for an award of SRF.